\documentclass[12pt]{article}

\usepackage{a4}
\usepackage{latexsym} 
\usepackage{amssymb}  
\usepackage{graphicx}
\usepackage{epsfig}
\usepackage{pslatex}
\usepackage{hyperref}

\def\spk#1#2{#1\, ({\it #2})}

\newenvironment{wparticipants}{\noindent {\sf PARTICIPANTS:}\\[2mm]}{\\[5mm]}

\newfont{\Thuge}{cmr10 scaled 8000}
\newfont{\thuge}{cmr10 scaled 6000}

\newcommand{\sqrtsnn}{\sqrt{s_{_{NN}}}}

\newcommand{\jpsi}{J/\psi}

\newcommand{\gaga}{\gamma\,\gamma}
\newcommand{\gp}{\gamma\,p}
\newcommand{\gA}{\gamma\,A}

\providecommand{\mumu}{\mu^+\mu^-}
\providecommand{\elel}{e^+e^-}
\providecommand{\lele}{l^{+}\,l^{-}}
\providecommand{\jpsi}{J/\psi}
\providecommand{\ups}{\Upsilon}

\providecommand{\STR}{{\sc starlight }}

\def\PRL{{\it Phys. Rev. Lett.\ }}

\def\NPA{{{\it Nucl. Phys.}~{\bf A}}}
\def\ttt#1{\texttt{\small #1}}

\newcounter{zyxabstract}     
\newcounter{zyxrefers}        

\newcommand{\newabstract}
{\newpage\stepcounter{zyxabstract}\setcounter{equation}{0}
\setcounter{footnote}{1}}

\newcommand{\rlabel}[1]{\label{zyx\arabic{zyxabstract}#1}}
\newcommand{\rref}[1]{\ref{zyx\arabic{zyxabstract}#1}}

\renewenvironment{thebibliography}[1] 
{\section*{References}\setcounter{zyxrefers}{0}
\begin{list}
{[\arabic{zyxrefers}]}
{\usecounter{zyxrefers}\setlength{\parindent}{0cm}\setlength{\itemsep}{0cm}}

}
{\end{list}}
{\section*{References}\setcounter{zyxrefers}{0}
\begin{list}{[\arabic{zyxrefers}]}
{\usecounter{zyxrefers}\setlength{\parindent}{0cm}\setlength{\itemsep}{-1.5mm}}}
{\end{list}}

\renewcommand{\bibitem}[1]{\item\rlabel{y#1}}
\renewcommand{\cite}[1]{[\rref{y#1}]}      

\begin{document}

\begin{center}
{\Large\bf Photoproduction at collider energies:} \\[0.2cm]
{\Large\bf from RHIC and HERA to the LHC}
\\[0.3cm]
{\large ECT*  -- Workshop, Trento,  January 15 - 19, 2007}\\[0.3cm]
A.~Baltz$^{1}$, G.~Baur$^{2}$, S.J.~Brodsky$^{3}$, ~D.~d'Enterria$^{4}$, 
U.~Dreyer$^{5}$, R.~Engel$^{6}$, L.~Frankfurt$^{7}$, Y.~Gorbunov$^{8}$, V.~Guzey$^{9}$,
A.~Hamilton$^{10}$, M.~Klasen$^{11}$, S.~R.~Klein$^{12}$, H.~Kowalski$^{13}$,
S.~Levonian$^{13}$, C.~Lourenco$^{4}$, M.V.T.~Machado$^{14}$, O.~Nachtmann$^{15}$,
Z.~Nagy$^{16}$, J.~Nystrand$^{17}$, K.~Piotrzkowski$^{18}$, P.~Ramalhete$^{19,4}$,
A.~Savin$^{20}$, E.~Scapparone$^{21}$, R.~Schicker$^{22}$, D.~Silvermyr$^{23}$, 
M.~Strikman$^{24}$, A.~Valkarova$^{25}$, R.~Vogt$^{12,26}$, and M.~Yilmaz$^{27}$\\[0.5cm]

{\it $^{1}$Brookhaven National Laboratory, Upton, NY 11973}\\
{\it $^{2}$ Institut f\"ur Kernphysik, Forschungszentrum J\"ulich, D-52425 J\"ulich}\\
{\it $^{3}$Stanford Linear Accelerator Center (SLAC), Stanford University, CA 94309}\\
{\it $^{4}$CERN, PH/EP, CH-1211 Geneva 23}\\
{\it $^{5}$Institute of Physics, University of Basel, CH-4056 Basel}\\
{\it $^{6}$Forschungszentrum Karlsruhe, Karlsruhe Institute of Technology}\\
{\it $^{7}$Department of Physics and Astronomy, Tel Aviv University}\\
{\it $^{8}$Creighton University, Omaha, 68178 NE}\\
{\it $^{9}$Ruhr U. Bochum, Inst. Theor. Physik II, D-44801 Bochum}\\
{\it $^{10}$Universit\'e de Gen\`eve, Facult\'e des Sciences, CH-1211 Geneva 4}\\
{\it $^{11}$Universit\'e Grenoble I, LPSC, F-38026 Grenoble Cedex}\\
{\it $^{12}$ Nuclear Science Division, LBNL, 1 Cyclotron Road, Berkeley CA 94720}\\
{\it $^{13}$DESY, D-22603 Hamburg}\\
{\it $^{14}$Centro de Ci\^encias Exatas e Tecnol\'ogicas, U. Fed. do Pampa, Bag\'e, RS}\\
{\it $^{15}$Institut f\"ur Theoretische Physik, Universit\"at Heidelberg, D-69120 Heidelberg}\\
{\it $^{16}$CERN, PH/TH, CH-1211 Geneva 23}\\
{\it $^{17}$Dept. of Physics \& Technology, University of Bergen, Allegaten 55, N-5007 Bergen}\\
{\it $^{18}$UC Louvain, B-1348 Louvain-la-Neuve}\\
{\it $^{19}$Instituto Superior Tecnico, Departamento de Fisica, 1049-001, Lisboa}\\
{\it $^{20}$Univ. of Wisconsin, Madison, WI 53706}\\
{\it $^{21}$Universit\'a di Bologna, I-40126 Bologna}\\
{\it $^{22}$Physikalisches Institut, Universit\"at Heidelberg, D-69120 Heidelberg}\\
{\it $^{23}$Oak Ridge National Laboratory, Oak Ridge, TN 37831}\\
{\it $^{24}$ Dept. of Physics, Pennsylvania State University, University Park, PA 16802}\\
{\it $^{25}$IPNP \& Charles University, CZ-180 00 Praha}\\
{\it $^{26}$University of California, Davis, CA 95616}\\
{\it $^{27}$Istanbul Tech U., TR-80626 Istanbul}\\

\vspace{0.5cm}
{\large ABSTRACT}
\end{center}
We present the mini-proceedings of the workshop on ``Photoproduction at collider energies: 
from RHIC and HERA to the LHC'' held at the European Centre for Theoretical Studies in 
Nuclear Physics and Related Areas (ECT*, Trento) from January 15 to 19, 2007. The workshop 
gathered both theorists and experimentalists to discuss the current status of investigations of 
high-energy photon-induced processes at different colliders (HERA, RHIC, and Tevatron) as well as 
preparations for extension of these studies at the LHC. The main physics topics covered were: 
(i) small-$x$ QCD in photoproduction studies with protons and in electromagnetic (aka. 
ultraperipheral) nucleus-nucleus collisions, (ii) hard diffraction physics at hadron colliders, 
and (iii) photon-photon collisions at very high energies: electroweak and beyond the Standard 
Model processes. These mini-proceedings consist of an introduction and short summaries of the talks 
presented at the meeting.\\\\
\vspace{0.4cm}
\noindent\rule{6cm}{0.3pt}\\


\section*{Introduction}

\noindent
Photon-induced collisions at high-energy are a fruitful tool to investigate strong and electromagnetic 
interactions. On the one hand, photon-hadron processes at the HERA $ep$ collider have provided precise 
information on QCD -- parton structure and evolution at low-$x$, partonic structure of the photon, 
measurement of the strong coupling constant, factorization breaking in diffractive processes, etc. -- 
via measurements of inclusive hard photoproduction (heavy-flavour, (di)jets, prompt photons ...) 
and exclusive diffractive vector meson production. 
On the other hand, ultra-relativistic proton and heavy-ion beams accelerated at BNL-RHIC ($\sqrtsnn$ = 200 GeV), 
FNAL-Tevatron ($\sqrt{s}$ = 1.96 TeV) and CERN-LHC ($\sqrtsnn$ = 5.5 -- 14 TeV) energies
generate strong electromagnetic fields which are equivalent to a flux of quasi-real photons 
with maximum energies in the range $E_{\gamma}^{max}\approx$~3 -- 2000 GeV. 
At the LHC, these photons offer the opportunity to study $\gp$, $\gA$ and $\gaga$ 
processes at energies one order of magnitude larger than at previous colliders, probing parton 
distributions at still lower $x$ values, and also opening interesting windows to electroweak 
and beyond the Standard Model physics.\\

\noindent
At heavy-ion colliders, the electromagnetic field due to the coherent action of the $Z\approx$ 80 
proton charges (for lead or gold nuclei) results in photon beams fluxes $Z^2\approx$ 7000 
times larger than those of corresponding electron or proton beams at the same energies. 
Photonuclear production of $\rho$ and $\jpsi$, as well as $e^+e^-$ pair production in
two-photon collisions have been studied in ``ultraperipheral collisions'' (UPCs) of gold nuclei
at RHIC. Studies of the small-$x$ QCD regime accessible in $\gA$ processes with UPCs at LHC are of 
critical importance for the interpretation of heavy ion data, and provide a clean tool for the study 
of non-linear QCD effects in the nuclear wave-function. In addition, coherent $\gaga$ processes 
in UPCs at the LHC allow one to study QED in a very strong field regime ($Z\alpha_{\rm em}\approx$ 0.6) 
as well to measure the couplings of the $\gamma,Z$ and $W^\pm$  gauge bosons among themselves.\\

\noindent
After 6 years of rich physics operation at HERA-II and RHIC, and less than one year before 
the start of LHC, it seemed a timely moment to have a workshop, gathering both theorists
and experimentalists, to discuss the current status of investigations of photon-induced processes
in $ep$ at HERA and UPC ion-ion collisions at RHIC, and preparations for extension of these studies
in PbPb and pp at LHC. A meeting was organized at the ECT* (Trento) from January 15--19, 2007, 
with partial financial support from the center. The meeting had 35 participants whose names and 
institutes are listed below. There were 30 presentations of various lengths. Ample time was left for 
discussions after each talk. The talks and discussions were organized around the following main topics:
\begin{itemize}
\item Overview theoretical and experimental talks on photoproduction at RHIC, HERA, Tevatron 
and prospects for LHC.
\item Inclusive and exclusive photoproduction at HERA: dijets, heavy-quarks, prompt photons, 
and vector mesons.
\item Vector-meson and hard- photoproduction in UPC ion-ion at RHIC and LHC, with emphasis
on the high gluon density regime at small-$x$ in the nucleus.
\item Photoproduction in $pp,p\bar{p}$ collisions at Tevatron and LHC: QCD (quarkonia,
jets, top), electroweak (W, Higgs) and beyond the SM processes.
\item Two-photon processes: studies of QED in strong field regime in UPCs, $\gaga\rightarrow\lele$ 
as a luminometer at the LHC, triple and quartic gauge $WW\gamma\,(\gamma)$ couplings.
\end{itemize}

\noindent
The programme, list of participants and a one-page summary of each talk including a few 
relevant references follow. We felt that this was more appropriate a format than full-fledged 
proceedings. Most results are or will soon be published and available on the archives.  
This approach leads to speedy publication and avoids too much duplication. 
Most of the talks can also be downloaded from the workshop website:\\

\centerline{ \ttt{http://cern.ch/david.denterria/photoprod\_ect07/}}

\bigskip

\noindent
We thank the ECT* management and secretariat, in particular Serena Degli Avancini and 
Cristina Costa, for the excellent organization of the workshop and all participants for their 
valuable contributions. We believe that this was only the first workshop of this kind and look 
forward to similar meetings in the future.\\

\vspace{0.6cm}

\noindent
{\sc Gerhard Baur,  David d'Enterria, Spencer Klein,}\\
{\sc Joakim Nystrand}, and {\sc Mark Strikman}

\vfill \eject

\section*{Programme}
\vspace{0.cm}

\begin{tabbing}
xx:xx \= A very very very long name \= \kill
{\bf Monday, 15 January 2007}\\

09:30 \hspace{0.4cm} \> {\it Open problems of small $x$ physics and UPC of heavy-ions} (1h00')  \> \hspace{7.2cm} L. Frankfurt  \\
11:00 \hspace{0.4cm} \> {\it On the non-perturbative foundations of the dipole picture} (1h00')   \> \hspace{7.2cm} O. Nachtmann  \\
14:00 \hspace{0.4cm} \> {\it Probing small $x$ dynamics in hard VM production phenomena} (50')   \> \hspace{7.2cm} M. Strikman  \\
14:50 \hspace{0.4cm} \> {\it Overview of photoproduction in CDF} (50')  \> \hspace{7.2cm}  A. Hamilton \\\\

{\bf  Tuesday 16 January 2007 }\\
09:30 \hspace{0.4cm} \> {\it Hard photoproduction at HERA: theory} (50')   \> \hspace{7.2cm} M. Klasen  \\
10:40 \hspace{0.4cm} \> {\it Hard photoproduction at HERA: experiment} (50')   \> \hspace{7.2cm} S. Levonian  \\
11:30 \hspace{0.4cm} \> {\it Review talk on parton shower algorithms} (1h00')   \> \hspace{7.2cm} Z. Nagy  \\
14:00 \hspace{0.4cm} \> {\it Electromagnetic nucleus-nucleus collisions: from RHIC to LHC} (50')   \> \hspace{7.2cm} R. Vogt  \\
14:50 \hspace{0.4cm} \> {\it Exclusive $\rho^0$ photoproduction in UPC with STAR} (40')   \> \hspace{7.2cm} Y. Gorbunov  \\
15:30 \hspace{0.4cm} \> {\it $\jpsi$ and dilepton pairs in electromagnetic AuAu at 200 GeV} (40')    \> \hspace{7.2cm}  D. Silvermyr   \\
16:10 \hspace{0.4cm} \> {\it Exclusive diffraction at Tevatron and prospects for the LHC} (40')   \> \hspace{7.2cm} A. Hamilton  \\\\

{\bf  Wednesday 17 January 2007 }\\
09:30 \hspace{0.4cm} \> {\it Diffractive and Photonic Reactions in QCD} (50')   \> \hspace{7.2cm} S. J. Brodsky  \\
10:40 \hspace{0.4cm} \> {\it Diffraction results at HERA} (30')    \> \hspace{7.2cm}  A. Valkarova  \\
11:10 \hspace{0.4cm} \> {\it Onium photoprod. and models of diffractive exclusive production} (50')    \> \hspace{7.2cm} H. Kowalski  \\
12:00 \hspace{0.4cm} \> {\it Hard photoproduction in ZEUS at HERA} (30')    \> \hspace{7.2cm} A. Savin  \\
14:00 \hspace{0.4cm} \> {\it Hard diffraction in DIS and pA} (50')    \> \hspace{7.2cm} V. Guzey  \\
14:50 \hspace{0.4cm} \> {\it Single $W$ boson photo-production in pp and pA collisions} (30')   \> \hspace{7.2cm} U. Dreyer  \\
15:20 \hspace{0.4cm} \> {\it Anomalous gauge couplings in $\elel$ and $\gaga$ scatt. at high energy} (40')  \> \hspace{7.2cm}  O. Nachtmann  \\\\

{\bf Thursday 18 January 2007 }\\
09:30 \hspace{0.4cm} \> {\it MC models for $\gp$ and $\gA$ interactions} (50')   \> \hspace{7.2cm} R. Engel \\
10:40 \hspace{0.4cm} \> {\it Low-$x$ QCD via electromagnetic PbPb colls. at 5.5 TeV in CMS} (40')  \> \hspace{7.2cm} D. d'Enterria, A. Hees\\
12:00 \hspace{0.4cm} \> {\it Trigger capabilities of the ALICE TOF for UPCs} (30')  \> \hspace{7.2cm}  E. Scapparone \\
14:00 \hspace{0.4cm} \> {\it Strong electromagnetic fields in UPCs: multiphoton processes} (40')   \> \hspace{7.2cm} G. Baur  \\
14:40 \hspace{0.4cm} \> {\it A low multiplicity trigger for peripheral collisions in ALICE} (30')   \>   \hspace{7.2cm} R. Schicker  \\
15:10 \hspace{0.4cm} \> {\it NA60 experimental capabilities for ultraperipheral collisions} (25')    \> \hspace{7.2cm} C. Lourenco  \\
15:40 \hspace{0.4cm} \> {\it Dimuon production in ultra-peripheral In-In collisions in NA60} (30')\> \hspace{7.2cm} P. Ramalhete  \\\\

{\bf Friday 19 January 2007 }\\
09:30  \hspace{0.4cm} \> {\it UPC lepton pair production to all orders in $Z\alpha$} (50')   \> \hspace{7.2cm} A. Baltz  \\
10:40 \hspace{0.4cm} \> {\it High energy photon interactions at the LHC} (50')   \> \hspace{7.2cm} K. Piotrzkowski   \\
11:30 \hspace{0.4cm} \> {\it Anomalous single top photoproduction at the LHC} (30')   \> \hspace{7.2cm} K. Piotrzk. J. de Favereau\\
14:00 \hspace{0.4cm} \> {\it Photoproduction and low-$x$} (50')  \> \hspace{7.2cm}  M. Machado  \\
14:50 \hspace{0.4cm} \> {\it Coulomb corrections in $\lele$ production in UPCs} (30')   \> \hspace{7.2cm} M. Yilmaz  \\
15:20 \hspace{0.4cm} \> {\it Photon-photon and photon-nucleus collisions in ALICE at LHC} (30')   \> \hspace{7.2cm} J. Nystrand\\\\

\end{tabbing}

%

\section*{List of Participants}

\begin{wparticipants}
\parbox[t]{0.47\textwidth}{
\spk{L. Frankfurt}{Tel Aviv U.},\\
\spk{R. Engel}{IKP Karlsruhe},\\
\spk{S. J. Brodsky}{SLAC},\\
\spk{V. Guzey}{Ruhr U. Bochum},\\
\spk{H. Kowalski}{DESY},\\
\spk{A. Baltz}{BNL},\\
\spk{R. Vogt}{LBL and UC Davis},\\
\spk{M. Machado}{U. Fed. do Pampa},\\
\spk{U. Dreyer}{U. Basel},\\
\spk{M. Yilmaz}{Istanbul Tech U.},\\
\spk{O. Nachtmann}{Heidelberg U.},\\
\spk{Z. Nagy}{CERN},\\
\spk{G. Baur}{IKP Juelich},\\
\spk{M. Klasen}{LPSC Grenoble},\\
\spk{M. Strikman}{Penn State U.},\\
\spk{C. Ewerz}{ECT*},\\
\spk{K. Itakura}{KEK Tsukuba},\\
\spk{K. Tywoniuk}{U. Oslo},\\
\spk{Y. Mehtar-Tani}{LPT Orsay},\\
}
\parbox[t]{0.53\textwidth}{
\spk{K. Piotrzkowski}{UC Louvain, CMS},\\
\spk{A. Hamilton}{Geneve U., CDF},\\
\spk{S. Levonian}{DESY},\\
\spk{A. Savin}{Wisconsin U., ZEUS},\\
\spk{A. Valkarova}{IPNP \& Charles U., H1},\\
\spk{J. Butterworth}{Creighton U., STAR},\\
\spk{Y. Gorbunov}{Creighton U., STAR},\\
\spk{D. Silvermyr}{ORNL, PHENIX},\\
\spk{R. Schicker}{Heidelberg U., ALICE},\\
\spk{P. Ramalhete}{IST-Lisbon, NA60},\\
\spk{D. d'Enterria}{CERN, CMS},\\
\spk{J. Nystrand}{Bergen U., ALICE},\\
\spk{C. Lourenco}{CERN, NA60},\\
\spk{E. Scapparone}{U. Bologna, ALICE}\\
}
\end{wparticipants}

%
\newabstract 

\footnotesize{

\begin{center}
{\large\bf Challenges of small $x$ hard QCD }\\[0.5cm]
{\bf L. ~Frankfurt}$^1$\\[0.3cm]
$^1$ Department of Physics and Astronomy, Tel Aviv University, Israel\\
\end{center}

\noindent
Challenges, puzzles of small $x$ QCD as well as promising methods to resolve
them are explained.  Firstly we outline basic theoretical results
obtained by methods of pQCD including lack of significant ambiguity in the
pQCD calculations in the kinematics of LHC and their success in the 
explanation of existing experimental data obtained at FNAL and HERA. 
Theoretical and experimental restrictions on the region of applicability 
of existing  pQCD methods and necessity of new QCD regime of strong 
interaction with small coupling constant are discussed (for the review see 
~\cite{annual}). We show that probability conservation and related complete
absorption at high energies prevent unrestricted increase with energy of
amplitudes of hard  processes at fixed impact parameter. Using
pQCD calculation of cross section for the scattering of small dipole off a
hadron target we demonstrate existence of boundary for the applicability of
methods of pQCD. We demonstrate that problems with probability conservation
become important for the gluon distribution within a nucleon in the end of
kinematics  of HERA and more clearly in the kinematics of leading
jets at RHIC and in a wider kinematical region at LHC. Rapid onset of
new regime is the combination of large value of gluon distribution within
a nucleon- $xG_N(x,Q^2)$ at low scale which is due to spontaneously broken 
chiral symmetry and fast increase with energy of pQCD amplitudes. Thus 
there is urging practical problem to investigate possible impact of this 
regime on the LHC physics including new particle production.
       Account of complete absorption gives us powerful and legitimate
method of calculation of amplitudes of hard processes in a region of
sufficiently small x. We explain that structure functions of a hadron
target will rapidly increase with energy forever:
$F_2(x,Q^2) =cQ^2 \mu^2 ln^3(x_o/x)$~\cite{Frankfurt:2001db}.\\

\noindent
At infinite energies $c$ and $\mu$ (universal tail of impact parameter
distribution within hadron target) are the same for hadrons and nuclei.
Factor $Q^2$ demonstrates complete violation of Bjorken scaling,of two
dimensional conformal invariance. Total cross section of nucleon
photodisintegration should increase faster with energy than cross sections
of hadronic collisions $ \sigma(\gamma\,N \to X)=c\ln^3(s/s_o)$.
HT effects form significant part of cross section. Golden plate experiment
will discover that cross section of HT process: $\gA\to 2jet +A$ should
constitute nearly 50\%  of the total cross section~\cite{Frankfurt:2001db}.
     We explain variety of phenomena which follow from regime of complete
absorption.  One of key theoretical tools is the account of the causality - 
the S.Mandelstam cancellation of the eikonal diagrams recently generalized 
to pQCD. Similar cancellations follow from energy-momentum conservation
~\cite{Blok:2006ns}. Critical prediction of QCD is increase with energy of 
the scale of processes having large cross section. One of new effects 
signaling onset of new QCD regime is energy losses of energetic partons 
produced in hadron-hadron and in hadron-nucleus collisions. Recent STAR 
data on leading parton production in dA collisions found :suppression of 
leading jet, presence of recoil jet and negative correlation with 
centrality trigger. All these features are well consistent with 
significant (around 10\%) energy losses for leading partons 
~\cite{Frankfurt:2006sc}. In the kinematics of LHC significantly more 
evidences for new QCD regime is expected and considered in the talk.

%
\newabstract 

\begin{center}
{\large\bf On the dipole picture in photon induced reactions }\\[0.5cm]
{\bf O. Nachtmann }$^1$\\[0.3cm]
$^1$Institut f\"ur Theoretische Physik, Universit\"at Heidelberg\\
Philosophenweg 16, D-69120 Heidelberg, Germany\\
\end{center}

\noindent
The idea that a high energy photon-hadron reaction can be related to a hadron-hadron 
reaction goes back to the 1960s. In the modern view the photon is supposed to fluctuate 
into a colour dipole quark-antiquark state long before the interaction. This color dipole 
interacts then with the hadron. We give a general nonperturbative treatment of photon-induced 
hadronic reactions based on the functional integral. The quark skeleton diagrams which 
are leading at high energies are identified. These diagrams are then separated into parts 
describing the photon splitting into $q\bar q$ and the $q\bar q$ scattering. Renormalisation 
is done using Schwinger-Dyson equations. This leads to the introduction of a rescattering 
term. The assumptions needed to go from there to the usual colour dipole picture of photon 
induced reactions are spelled out in detail. It is shown that the usual colour dipole picture 
for deep inelastic electron-proton scattering leads to rigorous inequalities for ratios of 
structure functions. We find an upper bound for $R=F_L/F_T$ and upper and lower bounds 
for the ratios of structure functions $F_2$ at the same $\gamma^*p$ c.m. energy $W$ but 
different $Q^2$. The comparison with the data from HERA suggests that the range of 
validity of the colour-dipole model may be more restricted than commonly thought.\\

\noindent
The results described in the talk are based on work done by 
C.~Ewerz and O.~Nachtmann, see hep-ph/0404254, 0604087, 0611076.

%
\newabstract 

\begin{center}
{\large\bf Probing small $x$ dynamics in hard vector meson production}\\[0.5cm]
{\bf M. ~Strikman}$^1$\\[0.3cm]
$^1$ Department of Physics, Penn State University, University Park, PA  16802, USA\\
\end{center}

\noindent
Study of production of exclusive and semiexclusive production of $J/\psi$ and 
$\Upsilon$ production at LHC will allow to study  elastic and inelastic interactions 
of small dipoles protons and nuclei in the a kinematic range 30 GeV $\ge W_{\gamma N}\ge$ 1 TeV 
with . The  coherent channel provides an effective method to probe the leading twist hard QCD regimes of 
color transparency and perturbative color opacity as well as the onset of black disk regime (BDR) 
in the soft and hard QCD interactions~\cite{Frankfurt:2001db}-\cite{Frankfurt:2003qy}. 
Study of UPC pA collisions will allow to study both exclusive onium production off protons and 
nuclei at much larger energy interval than for AA collisions~\cite{Frankfurt:2006tp}.\\

\noindent
The use of the neutron tagging will allow to select photon interactions at higher energies away from 
zero rapidity and will test dynamics of inelastic interactions of small dipoles~\cite{Strikman:2005ze}.
We demonstrate that study of large t vector meson photoproduction with rapidity gaps in ultraperipheral 
proton-ion and ion-ion collisions at LHC would allow to investigate the energy dependence of cross 
section of elastic scattering of a small dipole off the parton over a wide range of energies 
$10^3 < s_{dipole - parton} < 10^6$ GeV$^2$ where this cross section is expected to change by 
a factor $\ge 20$. The accessible energy range exceeds the one reached at HERA by a factor of 10 
both in $\gp$ and in $\gA$ scattering. In addition, study of A-dependence of this process 
will allow to determine the t range where interaction of small dipoles gives the dominant contribution 
and to investigate effects of absorption for the propagation of ultrarelativistic small $q\bar q$ dipoles 
through the nuclear media and probe in a novel way onset of the BDR~\cite{Frankfurt:2006wg}.

%
\newabstract 

\begin{center}
{\large\bf Overview of Photoproduction at CDF}\\[0.5cm]
{\bf A. Hamilton}$^1$, on behalf of the CDF Collaboration\\[0.3cm]
$^1$University of Geneva, Switzerland\\
\end{center}

\noindent
The Collider Detector at Fermilab (CDF)~\cite{CDF} has performed a search for
exclusive photoproduction of lepton pairs in $p\bar{p}$ collision data
at \mbox{$\sqrt{s}=1.96$~TeV}.  We report an observation of exclusive
\mbox{$\gaga \rightarrow e^+e^-$}, and a status report on the
search for exclusive $\mu^+\mu^-$ pairs.  In an exclusive dilepton
process there are no particles produced other than the lepton pair and
the incident hadrons do not dissociate.  
Exclusive events have been proposed as a search channel for new
physics at the Large Hadron Collider (LHC)~\cite{newphyslhc}.  The
primary advantage for exclusive $\gaga$ processes in the search
for new physics at the LHC is that the measurement of the outgoing
protons' momenta can provide an additional method to calculate the
invariant mass of the particles produced by the $\gaga$
system~\cite{fp420}.  Also, since the cross sections are generally
known with an accuracy better than $1\%$, exclusive dilepton processes
are interesting candidates for improving the typical $5\%$ uncertainty
on the luminosity measurement at hadron colliders~\cite{lhclumi}.\\

\noindent
The observation of exclusive \mbox{$\gaga \rightarrow e^+e^-$}
selects candidate events~\cite{exclee} from a data sample with ${\cal L} = 532 \pm
32$~pb$^{-1}$ by requiring an $e^+e^-$ pair ($E_T>5$~GeV and
$|\eta|<2.0$) with no other particles detected in CDF ($|\eta|<7.4$). The
proton and antiproton lose a small fraction of their momentum in these
collisions and escape along the beam direction without being detected.
The events are simulated using the {\sc lpair} Monte Carlo
generator~\cite{vermaseren}.
The selection efficiency for the candidate events is determined to be
$(1.6\pm0.2)\%$.  The efficiency is dominated by the requirement that
there be no other particles observed in the detector because bunch
crossings with an exclusive event as well as another $p\bar{p}$
interaction will be vetoed by this selection criteria.
A total of 16 candidate events pass the selection criteria.  Four
backgrounds are considered: jets that pass electron requirements (jet
fakes), cosmic rays that interact in the detector, non-exclusive
events, and \mbox{$\gaga \rightarrow e^+e^-$} events with
proton dissociation.  The jet fake background is estimated to be very
small, \mbox{$0.0^{+0.1}_{-0.0}$} events, and the cosmic background is
found to be negligible.  By fitting the inclusive $e^+e^-$ portion of
the data sample as a function of the number of clusters in the
calorimeter, the non-exclusive background is estimated to be
$0.3\pm0.1$ events.  The number of dissociation events in the 16 event
signal sample is estimated to be $1.6\pm0.3$ using both the {\sc
  grape-dilepton} and {\sc lpair} MC programs.
Therefore, the sum of all background sources is
\mbox{$N_{\mathrm{bkgd}}=1.9\pm0.3$} events.\\

\noindent
The cross section for exclusive \mbox{$p\bar{p} \rightarrow p + e^+e^-
  + \bar{p}$} is measured to be \mbox{$1.6
^{+0.5}_{-0.3}\mathrm{(stat)}\pm0.3\mathrm{(syst)}$~pb}.  This agrees
with the theoretical cross section \mbox{$1.71\pm0.01$~pb} predicted
by {\sc lpair}.  The kinematics of the events also agree with the
predictions of {\sc lpair}.  The probability of observing 16 events
when $1.9\pm0.3$ background events are expected is \mbox{$1.3
  \times10^{-9}$}, equivalent to a $5.5\sigma$ effect.
The search for exclusive \mbox{$p\bar{p} \rightarrow p + \mu^+\mu^- +
  \bar{p}$} has selected candidate events by requiring a $\mu^+\mu^-$
pair with $2.7<M_{\mu^+\mu^-}<4.0$ and no additional particles
observed in the detector.  Most of the candidate events have invariant
masses consistent with the $J/\psi$ or $\psi`$ resonances.  This
indicates that there could be significant contributions from
$\gamma$-Pomeron exchange.  The analysis is still on-going, with
results expected soon.

%
\newabstract 

\begin{center}
{\large\bf Hard photoproduction at HERA: theory}\\[0.5cm]
{\bf Michael Klasen}$^1$\\[0.3cm]
$^1$Universit\'e Grenoble I , Laboratoire de Physique Subatomique et de Cosmologie \\
F-38026 Grenoble Cedex, France\\
\end{center}

\noindent
We review the present theoretical understanding of photons and hard 
photoproduction processes at HERA, discussing the production of jets,
light and heavy hadrons, quarkonia, and prompt photons. Virtual and 
polarized polarized photons are also briefly discussed. The most important
leading and next-to-leading order quantum chromodynamics results are 
presented in analytic form, and a variety of numerical predictions 
is compared to recent data~\cite{klasen1}.\\

\noindent
We address in particular the extraction of the strong coupling constant 
from photon structure functions~\cite{klasen2} and inclusive jet measurements, the
infrared safety and computing time of jet definitions~\cite{klasen3}, the sensitivity 
of dijet cross sections on the parton densities in the photon, diffractive 
dijet photoproduction and the question of factorization breaking~\cite{klasen4}, the 
treatment of the heavy-quark mass in different variable flavor number 
schemes~\cite{klasen5}, the relevance of the color-octet mechanism to quarkonium 
production~\cite{klasen6}. We emphasize in particular the interplay of direct and 
resolved photon components in real and slightly virtual photon collisions.\\

\noindent
For dijet photoproduction in diffractive events~\cite{klasen4} and events with a 
leading neutron~\cite{klasen7}, we show that at next-to-leading order the predicted 
cross sections overshoot the data and that absorptive corrections are 
necessary. We propose a scale-independent factorization scheme of 
suppressing not only the resolved, but also the direct initial-state
contributions~\cite{klasen8}, and apply it to the transition region of virtual and
real photons~\cite{klasen9}. Within the Durham two-channel eikonal model~\cite{klasen10}, a 
reasonable description of H1 and ZEUS data with recent H1 diffractive 
parton densities (for diffraction) or GRV pion densities (for leading 
neutrons) can be achieved.

%
\newabstract 

\begin{center}
{\large\bf Hard photoproduction at HERA: experiment}\\[0.5cm]
{\bf Sergei Levonian}$^1$, on behalf of the H1 and ZEUS  Collaborations\\[0.3cm]
$^1$DESY, D-22603 Hamburg, Germany \\
\end{center}

\noindent
The largest cross sections in $ep$ collisions at HERA are due to quasi-real photons of low
virtuality, $Q^2 \approx 0$, interacting with protons. These photoproduction processes provide
rich possibilities to improve our understanding of the details of strong interactions at high 
energies by comparing experimental data with perturbative QCD calculations, which are presently
available up to ${\cal O}(\alpha_s^2)$.\\

\noindent
In this review a brief summary is given on jet production~\mbox{[1-9]}, prompt photons~[10-12]
and heavy quark final state~[13-17] studies at \mbox{HERA-1} by H1 and ZEUS collaborations.
The data are sensitive both to proton and photon structure in previously unexploited kinematical
range, as well as to the dynamics of hard scattering at partonic level.
For inclusive jets and dijet photoproduction NLO calculations provide fair description.
It is not yet the case for heavy quarks and multijet final states where clear indications are seen
for sizable higher order corrections, lacking in the theoretical calculations. The same conclusion
comes also from the fact, that for most of the measurements theoretical uncertainties dominate
over experimental errors.\\

\noindent
Scaled $E_T$ distributions~\cite{j1} demonstrate that inclusive jet spectra in $\gamma p$ interactions are 
harder than those in $p\bar{p}$ collisions. This can be attributed both to the direct photoproduction
and anomalous component of the photon structure.
$\alpha_s$ extracted from scaling violation in jet photoproduction~\cite{j2} with competitive precision agrees 
with world average value.Dijets were used to provide unique information on gluonic content of the photon down to 
$x_{\gamma}=0.05$~\cite{j3} as well as to constrain gluon in the proton at large $x_p$~\cite{j6},\cite{j7},\cite{Zf}.  
Angular distributions of dijets one of which containing $D^*$ is indicative of large fraction
of intrinsic charm in the photon (in LO framework)~\cite{hq2}.\\
 
\noindent
Results on the prompt photon and charm production show the sensitivity to the details of 
parton evolution. NLO calculations~\cite{LZ} using $k_T$-factorisation scheme and
unintegrated parton density functions provide best description of $\gamma+$jet cross sections.
Cascade~\cite{CAS} Monte Carlo program based upon CCFM evolution describes inclusive $D^*$
cross sections better than Monte Carlo programs based on DGLAP and collinear factorisation scheme.
Beauty photoproduction at HERA~[15-17], albeit statistically limited, provides additional valuable tests
of pQCD. Present NLO calculations tend to somewhat underestimate $b$ cross section in low $p_t$ range.
Many of those results will be further improved by using new \mbox{HERA-2} data, which amount to 3-fold
luminosity of HERA-1 run and were taken with upgraded detector capabilities.
In order to make best use of those data, we expect  corresponding progress from theory side.

%
\newabstract 

\begin{center}
{\large\bf Review talk on parton shower algorithms}\\[0.5cm]
{\bf Zoltan Nagy}$^1$\\[0.3cm]
$^1$ Physics Department, Theory Group CERN\\
CH-1211 Geneva 23, Switzerland\\
E-mail: {\tt Zoltan.Nagy@cern.ch}\\[0.5cm]
\end{center}

\noindent
In this talk I gave a short overview on the structure of the parton shower algorithms and 
the current development of the field. I compared the HERWIG~\cite{herwig}, 
PYTHIA~\cite{pythia} and ARIADNE~\cite{ariadne} models and I reviewed the matching problems.\\

\noindent
The largest contribution in high energy collisions comes from the hadronic event. The 
origin of the large hadronic contributions is the strong interaction, QCD. According
to the perturbation theory we are able to calculate the hard part (short distance physics part) 
of the cross section and deal with only few partons in the final state instead of branch of hadrons. 
But in the detector we can see lots of hadrons. The hadronization happens at low energy and we 
are not able to calculate the hadronization effect from the QCD but we can set up a suitable 
model for it. We can calculate the high energy part and we have model for the low energy part, 
so we need a {\em bridge} between these two parts. The bridge is the parton shower. Can we 
calculate high multiplicity cross section in the perturbative framework? We cannot do it exactly 
but approximately it is possible.\\

\noindent
The parton shower is based on the factorization property of the QCD matrix elements in the soft 
and collinear regions. In principle this would be a {\em well defined approximation} but in the 
parton shower algorithm we have further approximations top on the soft-collinear approximation: 
i) strong approximation in the phase space; ii) neglecting of the subleading color contribution 
($1/N_{c}^{2}$ expansion); iii) even the leading color correlations are approximated according 
to the angular ordering; iv) spin correlations are also neglected.\\

\noindent
These approximations allow us to have an algorithm that can produced unweighted hadronic events 
(as we can see in the detector) but we have to keep in mind that the parton shower is not the 
{\em nature}, it cannot be considered a QCD prediction. But we can improve them. One of the way 
is to add exact QCD matrix element. The CKKW~\cite{ckkw} algorithm deals with exact tree level 
matrix elements while the MC@NLO~\cite{mc@nlo} algorithm matches the shower calculations 
and the fix order NLO computations.

%
\newabstract 

\begin{center}
{\large\bf Determining the Nuclear Gluon Distribution in UPCs}\\[0.5cm]
{\bf R. Vogt}$^{1,2}$\\[0.3cm]
$^1$Nuclear Science Division, LBNL, Berkeley, CA, USA\\
$^2$Physics Department, University of California, Davis, CA, USA\\
\end{center}

\noindent
In this talk, we discuss photoproduction of hard processes, accessible at
RHIC and the LHC.  The high photon-nucleon center of mass energy, 
$\sqrt{s_{\gamma N}}$, and the large luminosities result in high rates for
$Q \overline Q$ (charm and bottom)~\cite{KNV}-\cite{SVW} and jet$+$jet~\cite{SVW} 
final states.  
In particular, at the LHC, the high energies and large rapidity coverage make
the nuclear gluon distribution accessible in the interesting low $x$ and 
perturbative $Q^2$ regime.  In AA collisions, both ions serve as photon
emitters and the effective photoproduction process is $\gA$. The
photon flux from a nucleus is enhanced by a factor of $Z^2$ relative to the
proton.  Thus in pA interactions, the $\gA$ luminosity, where the photon
comes from the proton, is negligible relative to the $\gp$ luminosity
where the the photon comes from the nucleus.\\

\noindent
We consider both direct photoproduction, where a photon from the
virtual photon field surrounding the nucleus interacts with a parton in the
opposite nucleus, and resolved production, where the virtual photon fluctuates
into states with multiple gluons and $q \overline q$ pairs, opening more
channels but increasing the average $x$ probed.  The nuclear gluon distribution
is more cleanly probed in direct photoproduction.  
The possible effects of nuclear modifications of the parton densities on the 
$\gA$ interactions, most important at low $x$ and moderate $Q^2$ 
~\cite{KNV}.  We compared two parameterizations of nuclear shadowing which
differ strongly for gluons, EKS98~\cite{EKS} and FGS~\cite{FGS}, to show the 
range of the measurable effect.  It is precisely this modification that these
LHC measurements can study.  The average $x$ values for the nucleon parton
momentum fraction are shown as a function of heavy quark $p_T$ (integrated
over rapidity) and rapidity ($p_T$ integrated) and jet $p_T$ (rapidity 
integrated).  We have also shown the $b \overline b$ and jet rates as a
histogram in $x$ and $p_T$ in the ATLAS acceptance~\cite{SVW}.\\

\noindent
Heavy flavor production is especially useful since, in direct photoproduction,
the only contribution is from $\gamma g \rightarrow Q \overline Q$.  Of course
heavy flavor measurements are rather difficult and jet measurements may be
simpler.  Jet production proceeds dominantly by $\gamma g \rightarrow q 
\overline q$ at low $p_T$ but includes a contribution from the QCD Compton
process, $\gamma q (\overline q) \rightarrow g q (\overline q)$ which, although
small at moderate $p_T$, increases with $p_T$.  The heavy flavor results are
shown as a function of quark $p_T$ and rapidity as well as $Q \overline Q$ pair
mass.  The jet results are shown as a function of jet $p_T$.  Final state
pion, kaon and proton $p_T$ distributions are also shown -- if jets are
not directly measurable, these leading hadrons may be.  
The results show that the rates are sufficiently high for these processes
to provide clean measurements of the nuclear gluon distribution for $\gA$
in AA and the proton parton distribution $\gp$ in pA interactions.

%
\newabstract 

\begin{center}
{\large\bf Photoproduction results in Ultra Peripheral\\ 
Relativistic Heavy Ion Collisions with STAR }\\[0.5cm]
{\bf Yury Gorbunov}$^1$, on behalf of the STAR Collaboration\\[0.3cm]
$^1$Creighton University, Omaha, NE 68178, USA\\
\end{center}

\noindent
We present a measurement of the coherent $\rho^{0}$ and direct $\pi^{+}\pi^{-}$
pair photo-production in ultra peripheral relativistic heavy ion
collisions at $\sqrt{s_{NN}}$=200 GeV. At impact parameters larger then twice the nuclear 
radius, the nuclei do not physically collide, but interact via long-range electromagnetic fields. 
The process $AuAu \rightarrow Au^{*}Au^{*}\rho^{0}$ with accompanying mutual nuclear excitations 
is observed. We report on $\rho^{0}$ production cross section with coherent and incoherent 
coupling accompanied by mutual nuclear break-up and on direct $\pi^{+}\pi^{-}$ production. The cross 
section has been studied as a function of $p_{T}$, $y_{rho^0}$ and $M_{\pi\pi}$.  The measured 
production cross section is compared to the theoretical models~\cite{kn}-\cite{fsz} and found to be in 
agreement with~\cite{kn}. $\rho^0$ helicity matrix elements have been measured and are consistent 
with S-channel helicity conservation.and in agreement with $\gp$ experiment~\cite{zeus}.\\

\noindent
We are also sensitive to the  interference between two production modes: the first nucleus emits a 
photon and scatters from the second or vice versa. We observe that $\rho^{0}$ production at low $p_{T}$ is 
suppressed, as expected if $\rho^{0}$ production at the two ions interfere destructively. 
The measured level of interference is 93 $\pm$ 6(stat.) $\pm$ 8(syst.) $\pm$ 15(theory)$\%$ of 
that expected~\cite{sklein}.  STAR has also studied $\rho^{0}$ production in $dAu$ collisions. 
Both coherent (deuteron stays intact) and incoherent (deuteron dissociates) interactions have been observed. 
The $p_{T}^{2}$ spectra matches those observed at eA at HERA~\cite{tim}.
The $AuAu \rightarrow \pi^{+}\pi^{-}\pi^{+}\pi^{-}$ state ($\rho^{0*}$ 1450) produced via 
excited states is observed and peaked at low $p_{T}<$ 100 MeV/c and the 4-pion mass spectra is peaked 
around 1.5 GeV/$c^2$. Finally,  lepton pairs production ( $e^+e^-$ ) accompanied by mutual 
Coulomb dissociation have been observed. The 
kinematic distribution match those predicted by the lowest order QED~\cite{epem}.

%
\newabstract 

\begin{center}
{\large\bf Photoproduction of $\jpsi$ and high mass $\elel$ in \\
Ultra-Peripheral Au+Au Collisions at $\sqrtsnn$~=~200~GeV}\\[0.5cm]
{\bf D. Silvermyr}$^1$, on behalf of the PHENIX Collaboration\\[0.3cm]
$^1$Oak Ridge National Laboratory, Oak Ridge, TN, USA\\\
\end{center}

\noindent
High energy ultra-peripheral collisions (UPC) of heavy-ions generate
strong electromagnetic fields which open the possibility to study 
$\gaga$ and $\gamma$-nucleus processes in a kinematic regime so far 
unexplored. In this talk, we discuss photoproduction of 
$\jpsi$ and high mass $\elel$ in Ultra-Peripheral 
Au+Au Collisions at $\sqrt{s_{NN}}$~=~200~GeV,
tagged with forward neutron emission from the $Au^\star$ dissociation. 
The electromagnetic field of any 
relativistic charged particle can be described as a flux of ``equivalent'' photons,
originally due to Fermi~\cite{Fermi:1925fq}, and given by the 
Weizs\"acker-Williams~\cite{vonWeizsacker:1934sx} 
or Equivalent Photon Approximation (EPA) formula.
The validity of the application of the EPA formula for heavy-ions 
is limited to the case where {\it all} the protons of the nucleus interact 
electromagnetically in a coherent way. In that case, the wavelength of the resulting 
photon is larger than the size of the nucleus, given by its radius $R_A$.
This ``coherence'' condition limits the maximum virtuality of the produced photon 
to very low values~\cite{Baur:1998ay}. 
The two topics discussed in this talk are the coherent photoproduction of:
\begin{description}
\item I. $\jpsi$, the heaviest vector meson effectively accessible in $\gA$ 
collisions at RHIC, via: 
$A+A \;(\rightarrow \gamma+A) \rightarrow A^\star+A^{(\star)}+\jpsi$.
\item II. High mass di-electron continuum in $\gaga$ collisions:
$A+A \; (\rightarrow \gamma+\gamma) \rightarrow A^\star+A^{(\star)}+e^++e^-$.
\end{description}
For the first process (I), the cross-section for heavy vector 
meson ($\jpsi,\Upsilon$) photoproduction is found to depend 
(i) {\it quadratically} on the gluon density $G_A(x,Q^2)$~\cite{Ryskin:1995hz}
as well as on (ii) the probability of rescattering or absorption of the $Q\overline{Q}$ 
pair as it traverses the nucleus. The study of quarkonia production in 
$\gA$ collisions at RHIC or LHC energies is, thus, considered a good
probe of (i) the gluon distribution function $G_A(x,Q^2)$ in nuclei, and
(ii) vector-meson dynamics in nuclear matter.\\

\noindent
The PHENIX data taking during Run4 (2004) included a special UPC trigger requiring
a charged particle veto rapidity gap (empty $|y| \in [3,4]$ region), 
at least one single particle (cluster) seen in the mid-rapidity calorimeter with an energy of
at least 800 MeV, and that the zero-degree calorimeters should see at least one 
neutron emitted at very forward rapidities. 
Roughly 40\% of the UPCs leading to $\jpsi$ production, fulfill this latter requirement due 
to Coulomb excitation of at least one of the two nuclei.
After event selection, offline analysis cuts (incl. electron ID), and like-sign
combinatorial background subtraction, a few tens of di-electron pairs remain in the data
sample. 
These pairs can then be further divided up into coherent and incoherent $\jpsi$ candidates, 
and continuum dielectron pairs, via their invariant mass ($M_{inv}$) and transverse 
momentum ($p_{T}$) distributions. 
The preliminary PHENIX results~\cite{d'Enterria:2006ep} on these topics
are presented, illustrating that the trigger and analysis scheme is functional.
The current analysis is however clearly statistically limited, but with future higher luminosity runs 
(expected in 2007-08) more significant measurements should be possible.

%
\newabstract 

\begin{center}
{\large\bf Exclusive Diffraction at CDF}\\[0.5cm]
{\bf A. Hamilton}$^1$, on behalf of the CDF Collaboration\\[0.3cm]
$^1$University of Geneva, Switzerland\\
\end{center}

\noindent
Diffractive pp collisions are mediated by the exchange of a
colorless object with the quantum numbers of the vacuum, the Pomeron,
which allows the protons to stay intact while still producing some
central system $X$, $pp \to p + X + p$.  If $X$ is completely
determined, then the interaction is called an {\it exclusive
  diffractive} collision.  Exclusive diffractive processes heavily
favour final states with the quantum number of the vacuum,
\mbox{$J^{PC}=0^{++}$}. 
Exclusive diffraction could play a significant role in the observation
(and precision measurement) of the Standard Model (SM) Higgs boson and
other beyond the SM processes at the LHC - if suitable proton tagging
detectors are installed~\cite{cox}.  The advantages of exclusive
processes rely on two things: 1) $M_X$ can be determined by the
momentum of the outgoing protons, 2) $X$ is heavily favoured to have
\mbox{$J^{PC}=0^{++}$}.  The detectors necessary for proton tagging at
ATLAS and CMS are being developed by the FP420
Collaboration~\cite{fp420}
One of the challenges in the development of the exclusive experimental
program at LHC, is the lack of confirmation of the theoretical models
for exclusive diffraction.  Two theoretical models have been proposed
and developed to the point where they can be experimentally tested;
the Durham model~\cite{durham} implemented in the {\sc ExHuME}
MC~\cite{exhume} and the Saclay model~\cite{saclay} implemented in the
{\sc DPEMC} MC~\cite{dpemc}.

\noindent
The Collider Detector at Fermilab (CDF)~\cite{CDF} is attempting to distinguish
these models using $p\bar{p}$ collisions at
\mbox{$\sqrt{s}=1.96$~TeV}.  CDF has looked at three exclusive
diffractive production processes; $\gaga$ (through a quark
loop), dijet, and $\chi_c \to J\psi+\gamma$.
In the $\gaga$ channel three candidate events have been
observed, but the background estimates are not yet complete.  The {\sc
  ExHuME} MC predicts $1^{+3}_{-1}$ events expected in the data sample
examined in the $\gaga$ analysis.
Because the definition of an {\it exclusive jet} is experimentally
ambiguous, the dijet channel examines a variable called the {\it dijet
  mass fraction}, $R_{jj}\equiv\frac{M_{jj}}{M_{X}}$, where $M_{jj}$
is the mass of the dijet system and $M_{X}$ is the mass reconstructed
using all the observed particles in the detector.  The dijet channel
has extracted an exclusive dijet signal in the high $R_{jj}$ region of
an inclusive diffractive dijet sample.  The signal favors the {\sc
  ExHuME} MC predictions over the {\sc DPEMC} predictions.  By using
the fraction of heavy flavor jets in the inclusive diffractive dijet
sample, the \mbox{$J^{PC}=0^{++}$} selection of exclusive events has
also been observed.\\

\noindent
In the $\chi_c$ channel, candidate events have been selected, but
detailed analysis of acceptance, efficiency, and backgrounds are not
yet complete.  Backgrounds from $\gamma$-Pomeron interactions to
$J\psi$ are expected.
In conclusion, the CDF Collaboration is actively searching for
exclusive diffractive processes to distinguish proposed theoretical
models.  The results in the dijet channel favour the Durham model of
exclusive production over the Saclay model.

%
\newabstract 

\begin{center}
{\large\bf Photonic and Diffractive Processes in QCD}\\[0.5cm]
{\bf Stanley J. Brodsky}$^1$\\[0.3cm]
$^1$Stanford Linear Accelerator Center, Stanford University, California 94309\\
\end{center}

\noindent
The prospect of proton-proton and ion-ion collisions at the LHC
collider has led to a new focus on diffractive collisions where one
or both of the projectiles remain intact.  In this talk I emphasize
a number of novel physics features of photon-photon and
photon-Pomeron collisions which are accessible in diffractive and
ultra-peripheral reactions.\\

\noindent
1. The equivalent photon distribution~\cite{Brodsky:1971ud} of a
nucleus in  light-cone fraction $x$ and $k_\perp$ is most easily
derived using frame-independent light-front methods. Nuclear
coherence is maintained for $x <  (M_A R_A)^{-1},$ leading to a
remarkably large kinematic range for particle production from
photon-photon and photon-Pomeron interactions up to hundreds of GeV
for heavy ion collisions at the LHC.  Coulomb corrections as
expressed by the Schwinger-Sommerfeld
interaction~\cite{Brodsky:1995ds} with large nuclear charge give
large distortions of the trajectories of the  charged particles,
particularly when relative velocities are small.  The Coulomb
corrections also produce interesting charge asymmetries in
ultra-peripheral lepton-pair production~\cite{Brodsky:1968rd}.
Hadron production in diffractive interactions can involve both
photon and multi-gluon/Pomeron exchange, giving amplitudes with
different charge conjugation which can interfere and thus also
produce charge asymmetries.\\

\noindent
2. The elastic Coulomb scattering of heavy ions is modified by
quantum corrections associated with vacuum polarization and
light-by-light scattering, giving ${\cal O}(\alpha$) corrections to
the effective Lippmann-Schwinger kernel when $Z \alpha  = {\cal
O}(1)$.\\

\noindent
3. Diffractive deep inelastic scattering $\ell p \to \ell  p + X$
arises in QCD at leading twist from the final-state interactions of
the struck quark~\cite{Brodsky:2002ue}.   This in turn leads to
nuclear shadowing and corrections to the observed structure
functions of hadrons and nuclei at leading twist which are not in
the wavefunction of the target hadron computed in
isolation~\cite{Brodsky:2004qa}.  Such final-state interactions also
give single-spin asymmetries in single-inclusive DIS which measure
the orbital angular momentum of quarks and gluons in the target
hadron~\cite{Brodsky:2002cx}-\cite{Lu:2006kt}.  In the case of the
Drell-Yan reaction, one gets analogous leading-twist effects from
initial-state interactions, but with a sign reversal of the
single-spin asymmetry~\cite{Collins:2002kn}-\cite{Brodsky:2002rv}. When
one allows for both the quark and antiquark to interact  in the
initial state, one gets a leading-twist contribution to the $\cos
2\phi$ distribution of the lepton pair which violates the Lam-Tung
relation of PQCD~\cite{Boer:2002ju}-\cite{Boer:2005en}.  An important
lesson is that one must include - even at leading twist - the
interactions of the active quarks with the spectator quarks due to
the ``dangling gluons" associated with the Wilson line. The same
effects arise from other Feynman diagrams in the case of light-cone
gauge~\cite{Brodsky:2002ue}.\\

\noindent
4. Diffractive dijet production, such as $\pi A \to {\rm Jet Jet} +
A$ as measured in the E791 experiment at
Fermilab~\cite{Ashery:2005wa} not only tests QCD color
transparency~\cite{Brodsky:1988xz}, a fundamental property of color
gauge theory, but they also measure the frame-independent
light-front wavefunction of the projectile~\cite{Frankfurt:2000jm} .
Such measurements can be used to determine the proton LFWF by
measuring $e p \to e {\rm Jet Jet Jet}$ at HERA or diffractive
tri-jet production $ p A \to  {\rm Jet Jet Jet} + A$ in proton-ion
collisions.  The hadronic input, the nonperturbative LFWFs of
hadrons can be predicted using AdS/CFT~\cite{Brodsky:2006uq} .
Diffractive $d A$ reactions can also provide a new test of ``hidden
color" in the deuteron wavefunction~\cite{Brodsky:1983vf}. Hard
exclusive processes, such as $\gamma \gamma \to H \bar H$ at large
$s$ and $-t$, and diffractive photoproduction $\gp \to V^0 p$
are also sensitive to the shape of the hadron light-front
wavefunction or distribution amplitude of the produced
mesons~\cite{Brodsky:1994kf}.\\

\noindent
5. The existence of intrinsic charm and bottom Fock
states~\cite{Brodsky:1980pb} in the proton  $|uud Q \bar Q>$ with
probability $1/M^2_Q$~\cite{Franz:2000ee} leads to doubly
diffractive processes such as $p p \to \Upsilon + p+ p$ and  $p p
\to H + p+ p$~\cite{Brodsky:2006wb}, where a heavy quarkonium state
such as an $\Upsilon$ or the neutral Higgs couples to both members
of the intrinsic heavy quark pair and thus can attain a longitudinal
momentum fraction $x_F$ as large as $0.8$.  The fact that the
intrinsic heavy quark Fock states have a large color-octet dipole
moment, such as $|(uud)_{8C} (c \bar c)_{8C}>$  requires that the
gluon exchange producing the color-singlet quarkonium  occurs at the
nuclear surface, thus explaining the remarkable $A^{2/3}$ behavior
of $p A \to J/\psi X$ cross section observed at high
$x_F$~\cite{Leitch:2006ff}. Double intrinsic charm Fock
states~\cite{Vogt:1995tf} lead to double $J/\psi$ production as
observed by NA3~\cite{Badier:1982ae} and the doubly charmed baryons
observed by SELEX~\cite{Ocherashvili:2004hi} at large $x_F.$\\

\noindent
{\bf Acknowledgments}
\noindent
Work supported by the Department of Energy under
contract number DE--AC02--76SF00515.

%
\newabstract 

\begin{center}
{\large\bf Diffraction at HERA}\footnote{This work is supported by the Center for Particle Physics, project No. LN 00A006.}\\[0.5cm]
{\bf Alice Valk\'arov\'a}$^1$, on behalf of the H1 and ZEUS  Collaborations\\[0.3cm]
$^1$Institute of Particle and Nuclear Physics of the Charles University,\\ 
V Hole\v sovi\v ck\'ach 2, Praha 8, 180 00, Czech Republic\\
\end{center}

\noindent
{\bf Factorization in diffraction:}\\
\noindent
The QCD properties of diffraction can be studied at HERA via diffractive positron 
proton interactions of type $e^+p \rightarrow e^+XY$. There hadronic diffractive 
system $x$ is separated by large rapidity gap (due to colour singlet exchange) from the 
system Y  which consists of a proton or a low mass system.The QCD collinear factorization 
implies that diffractive processes in deep inelastic scattering (DDIS) can be understood as 
a convolution of the process dependent perturbatively calculable matrix elements with 
universal diffractive parton density functions (DPDF), which gather the nonperturbative 
components of diffractive process~\cite{alice1}. These are determined from NLO QCD fits to 
measurements of inclusive diffractive DIS cross sections within the factorizable Pomeron 
model and using DGLAP evolution equations~\cite{alice2}. If QCD factorization is fulfilled, 
NLO QCD calculation based on these DPDFs should be able to predict the production 
rates of  exclusive diffractive processes like dijet and open charm production. 
It has been found that in the regime of DIS the predictions of QCD are in good agreement 
with the experimental results~\cite{alice3}-\cite{alice6}.
In the regime of photoproduction the cross section for dijet production was however found 
to be suppressed by factor about 1.6 in comparison with NLO QCD predictions thus 
indicating  QCD factorization breaking in photoproduction~\cite{alice7}-\cite{alice8}. On the contrary 
there is not observed the evidence for such a suppression (within large theoretical and 
experimental errors) in open charm photoproduction~\cite{alice5}-\cite{alice9}.\\

\noindent
{\bf Vector meson production:}\\
\noindent
>From the measurement of the $W$ dependence of the exclusive $\rho^0$ photoproduction  
the Pomeron trajectory $\alpha_P(t)$ has been extracted. The value $\alpha^{\prime}$ 
was found to be smaller than the value 0.25 $\boldmath {GeV^2}$~\cite{alice10}.
The cross section for elastic J/$\psi$ in photoproduction was measured~\cite{alice11}. 
Within models, the data show a high sensitivity to the gluon density of the proton 
in the low $x$ and low $Q^2$ region. 
The diffractive photoproduction of $\rho^0$~\cite{alice12} and  J/$\psi$~\cite{alice13}-\cite{alice14} 
with large momentum squared at the proton vertex, $|t|$, has been investigated.
The data of diffractive photoproduction of $\rho^0$
indicate a violation of s-channel helicity conservation\cite{alice12}, on the contrary 
J/$\psi$ production data were found to be compatible with helicity conservation~\cite{alice14}. 
Data were compared to the predictions of QCD perturbative models.

%
\newabstract 

\begin{center}
{\large\bf Onium photoproduction and \\ models of diffractive exclusive production}\\[0.5cm]
{\bf Henri Kowalski}$^1$\\[0.3cm]
$^1$DESY, D-22603 Hamburg\\
\end{center}

\noindent
One of the most important discovery of HERA was the observation of abundant 
diffractive reactions, in which the proton remained intact and lost only a small fraction 
of its momentum. The fraction of diffractive processes is well above 10\% at 
$Q^2=10$ GeV$^2$ and decreases only logarithmically with increasing $Q^2$. 
The abundance of diffractive reactions indicates that they are the shadow of inclusive 
DIS processes. At low $x$, the leading order QCD description of DIS (in the momentum 
space) is equivalent to the scattering of small dipoles (defined in the configuration space) 
and the optical theorem relates directly inclusive and diffractive scattering.\\

\noindent
For small dipoles, with a size $r$, the dipole cross section is proportional to the gluon 
density and $r^2$,  
\begin{equation}
\label{eq:sigdip}
\sigma^{\gamma^* p}(r^2) \propto r^2\,\alpha_S\, xg(x,Q^2)
\end{equation}
The inclusive DIS cross section is proportional to the dipole cross section whereas the 
diffractive ones are proportional to its square.\\

\noindent
In the dipole picture  the physical cross section is obtained by averaging the dipole 
cross section with probabilities to find a dipole of size $r$ given by the wave functions.   
For the inclusive DIS process only the photon wave function is needed and it is known 
from QED.   The only unknown quantity is the gluon density, and it is then extracted 
from fits to data.  For vector meson production, the vector meson wavefunction is also 
needed, and there exist good approximations. 

\noindent
The  dipole models provide a simultaneous description of $F_2$ and inclusive and 
exclusive diffractive measurements at HERA of astonishing quality~\cite{Bartels:2002cj}-\cite{Kowalski:2006}. 
The cross section for exclusive $J/\Psi, \, \phi, \, \rho$ photo- and electro-production 
and DVCS process and the ratio of the cross sections for longitudinally and transversely 
polarized vector mesons are very well described as a function of $Q^2,\, W^2$ and 
the squared momentum transfer, $t$. One of the results of this type of analysis is a clear 
indication of saturation effects in the core of the proton~\cite{Kowalski:2003hm}. 

  
%

%
\newabstract 

\begin{center}
{\large\bf Photoproduction of Jets with ZEUS at HERA}\\[0.5cm]
{\bf A.~A.~Savin}$^1$, on behalf of the ZEUS Collaboration\\[0.3cm]
$^1$University of Wisconsin,Madison,USA\\
\end{center}

\noindent
In photoproduction, a quasi­real photon, emitted by the incoming positron, interacts with a 
parton from the proton. Photoproduction may be categorised at leading order (LO) as being 
either direct, if the photon interacts as a point­like particle, or resolved, if it 
fluctuates into a partonic system, and subsequently transfers only a fraction of its 
momentum to the hard interaction. In resolved photoproduction more than one pair of 
partons from the incoming hadrons may interact, e.g. multi­parton interactions (MPIs) may 
take place. The secondary scatters generate additional hadronic energy flow in the event, 
the topology of which is poorly understood theoretically. \\

\noindent
Differential cross sections for three­ and four-jet photoproduction events in a wide 
region of jet-system invariant mass  $M_{3(4)j}$  demonstrate, that the resolved part of 
the cross sections at low values of $M_{3(4)j}$ can be only described by the models 
by adding the MPI process to the standard MC simulation. Therefore this region is of 
particular interest for studying of the MPI regime at HERA. The pQCD calculations, existing 
currently only at order of $O(\alpha, \alpha_S^2)$ describe the data at high 
values of $M_{3j}$, but do it much poorly in the low $M_{3j}$ region.\\
 
\noindent
The production of events with two high transverse energy jets in the final state 
separated by a large rapidity interval provides an ideal environment to study the 
interplay between soft (non­perturbative) and hard (perturbative) QCD. The dominant 
mechanism for the production of jets with high transverse energy in hadronic collisions is 
a hard interaction between partons in the incoming hadrons via a quark or gluon propagator. 
The exchange of color quantum numbers generally gives rise to jets in the final state that 
are color connected to each other and to the remnants of the incoming hadrons. 
This leads to energy flow populating the pseudorapidity region both between the jets 
and the hadronic remnants, and between the jets themselves. The fraction of events with 
little or no hadronic activity between the jets is expected to be exponentially suppressed 
as the rapidity interval between the jets increases. A non­exponentially suppressed 
fraction of such events would therefore be a signature of the exchange of a color­singlet 
object. The high transverse energy of the jets provides a perturbative hard scale at 
each end of  the color­singlet exchange, so that the cross section should be 
calculable in perturbative  QCD.\\

\noindent
Dijet photoproduction has been measured for configurations in which the two jets with 
highest transverse energy are separated by a large rapidity gap. The fraction of events 
with very little transverse energy between the jets is inconsistent with the predictions 
of standard photoproduction MC models. The same models with the inclusion of a color­singlet 
exchange sample at the level of 2-3\% are able to describe the data, including the 
gap­fraction dependency on different kinematic variables.

\providecommand{\etal}{et al.~}
\providecommand{\coll}{Collab.~}

%
\newabstract 

\begin{center}
{\large\bf Hard diffraction in DIS and pA}\\[0.5cm]
{\bf V. Guzey}$^1$\\[0.3cm]
$^1$Ruhr U. Bochum, Germany\\
\end{center}

\noindent
In this talk, we reviewed several coherent diffractive processes 
with nuclear targets. They included soft hadron-nucleus diffraction,
hard inclusive diffraction in Deep Inelastic Scattering (DIS) on
nuclear targets and hard diffraction in proton-nucleus scattering.
We discussed the leading twist theory of nuclear shadowing~\cite{guzey1}-\cite{guzey4}
and presented predictions for usual and diffractive nuclear parton
distribution functions~\cite{guzey5}.\\

\noindent
We presented a novel method for the calculation of hard coherent 
proton-nucleus diffraction~\cite{guzey6} using the leading twist theory of 
nuclear shadowing. We discussed the novel dramatic prediction 
that soft multiple interactions lead to very large factorization 
breaking in hard coherent pA diffraction.
We compared hard and e.m. mechanisms of hard coherent diffractive
production of two jets in pA scattering. We observed that:
(i) For heavy nuclei (Pb-208) at the LHC, hard diffraction is 
suppressed compared to the e.m. mechanism;
(ii) for lighter nuclei (Ca-40), hard diffraction is compatible to
the e.m. mechanism for the production of quark and
gluon jets; hard diffraction is suppressed compared to the 
e.m. mechanism for heavy-quark-jets.\\

\noindent
This suggests the following experimental strategies:
(i) The use of heavy nuclei would allow to cleanly study the $\gp$ interaction;
 (ii) The use of lighter nuclei would allow to study factorization 
   breaking in  nuclear diffractive PDFs;
 (iii) In the same kinematics, a comparison of diffractive dijet production to 
   heavy-quark-jet production would probe nuclear diffractive gluon PDF.\\

%
\newabstract 

\begin{center}
{\large\bf $W$~boson photoproduction in pp and pA collisions at LHC}\\[0.5cm]
{\bf Ute Dreyer}$^1$,  {\bf Kai Hencken}$^1$, and {\bf Dirk Trautmann}$^1$\\[0.3cm]
$^1$Institute of Physics, University of Basel, Klingelbergstr. 82, 4056 Basel, Switzerland\\
\end{center}

\noindent
The couplings of gauge bosons among themselves belong to one of the least tested 
sectors of the electroweak theory. A process well-suited to testing the gauge boson 
self-interaction is the photoproduction of single $W$~bosons from a nucleon. In particular, 
this process is sensitive to the $WW\gamma$~coupling. In the framework of the equivalent 
photon approximation~\cite{Fermi}-\cite{Baur:2001jj} either one 
of the protons in p-p collisions or the ion in p-A collisions at LHC is replaced by a spectrum of 
equivalent photons. Thus, the production rate of single $W$~bosons in ultraperipheral
collisions can be determined either from the convolution of the equivalent photon spectrum 
with the exclusive process $\gamma + p \rightarrow W^+ + n$
or the corresponding inclusive process. While the cross section for the
inclusive process is expected to be larger, the exclusive process has the 
advantage of a neutron in forward direction with approximately the energy of the initial proton.\\

\noindent
First, we calculate the cross section for real photoproduction of single $W$~bosons and cross-check 
our results with those of Fearing {\em et al.}~\cite{fearingphoton}-\cite{fearingmuon}, 
who have calculated this cross section with a smaller $W$~boson mass in mind. Next, 
we extend the calculations of Fearing {\em et al.} by including a weak magnetic form 
factor and the correct $W$~boson mass, and convolve the photoproduction cross sections 
with the equivalent photon spectra of ions and protons~\cite{Budnev}-\cite{Kniehl} 
in order to obtain results for p-p and p-A collisions.\\

\noindent
We show how the choice of the weak timelike form factors affects the sensitivity of the 
total cross section to the anomalous magnetic moment of the $W$~boson.
Furthermore, we give an estimate of the total cross sections for p-p ($\sim 10^{-39}cm^2$) 
and Pb-p ($\sim 10^{-36} cm^2$) collisions. However, the feasibility of measuring this 
process is still an open question. For p-p collisions yet another possibility exists. For large 
momentum transfers $Q^2$ of the photon, the proton should be regarded as a collection 
of partons which in turn can be approximated by an equivalent stream of 
photons~\cite{drees}-\cite{pisano}. Convolving the photoproduction cross sections 
with the deep inelastic equivalent photon spectrum over the appropriate range of momentum 
transfers, we obtain rates in a similar range as for the elastic case. Whether the dominant 
diagrams are the same as for real photoproduction is still to be checked.

%
\newabstract 

\begin{center}
{\large\bf Anomalous gauge boson couplings in \\
$e^-e^+\to W^-W^+$ and $\gaga \to W^-W^+$ at an ILC}\\[0.5cm]
{\bf O. Nachtmann}$^1$\\[0.3cm]
$^1$Institut f\"ur Theoretische Physik, Universit\"at Heidelberg\\
Philosophenweg 16, D-69120 Heidelberg, Germany\\
\end{center}

\noindent
One important physics topic which will be studied at a $e^+e^-$ linear collider, 
the ILC, with c.m. energies $0.5$ to $1.0$ TeV concerns the couplings of the gauge 
bosons $\gamma,Z$ and $W^\pm$ among themselves. In the Standard Model (SM) 
these couplings are fixed by the requirement of renormalisability. Any deviation from 
the SM couplings would, therefore, signal new physics. We describe the accuracies 
achievable for anomalous couplings in various proposed modes of operation of the ILC. 
The reactions $e^-e^+\to W^-W^+,\gaga\to W^-W^+$ and $Z$ 
decays at Giga $Z$ are considered. From the theoretical side we follow first a form 
factor approach. It is known that $28$ real parameters describe the general $\gamma WW$ 
and $ZWW$ vertices. All of these parameters can be measured in $e^-e^+\to W^-W^+$ 
at an ILC if also transversely polarised $e^-$ and $e^+$ beams are available. 
A particularly interesting option for the ILC is a $\gaga$ collider. To compare 
the sensitivities reachable for anomalous gauge boson couplings in $e^-e^+$ and 
$\gaga$ collisions we use an effective Lagrangian approach. It turns out that 
in such an approach also $Z$ decay observables become sensitive to anomalous gauge couplings. 
From presently available electroweak precision measurements we deduce a correlation 
between the value for the Higgs mass and a certain anomalous coupling. Already small 
values for the latter allow Higgs masses up to $500$ GeV. At an ILC the various reactions, 
$e^-e^+\to W^-W^+,\gaga\to W^-W^+$ and $Z$ decays give complementary 
information on the gauge-boson couplings. The indirect reach for new physics in these reactions 
will be up to around $10$ TeV.\\

\noindent
The results presented in the talk are based on work done by M.~Diehl, F.~Nagel, M.~Pospischil, 
A.~Utermann and O.~Nachtmann, see Z. Phys. {\bf C62}, 397 (1994) and 
Eur. Phys. J. {\bf C1}, 177 (1998), {\bf C27}, 375 (2003), {\bf C32}, 17 (2003), {\bf C40}, 497 (2005), 
{\bf C42}, 139 (2005), {\bf C45}, 679 (2006), and {\bf C46}, 93 (2006).

%
\newabstract 

\begin{center}
{\large\bf Simulation of Photoproduction on Nuclei and \\ Astroparticle Physics Connection}\\[0.5cm]
{\bf Ralph Engel }$^1$\\[0.3cm]
$^1$Forschungszentrum Karlsruhe, Karlsruhe Institute of Technology\\
\end{center}

\noindent
The interaction of photons with nuclei is an important process in
astroparticle physics. It takes place in the source regions where
cosmic rays are accelerated, leads to energy losses and secondary
particle production during propagation, and determines the muon
content of em. showers in the Earth's atmosphere. On the other hand,
photoproduction interactions provide an interesting means of
investigating QCD processes in accelerator experiments and also
represent background reactions for the investigation of rare
processes.\\

\noindent
In the presentation, the connection between astroparticle physics and
high energy physics is illustrated using the particularly striking
example of ultra-high energy cosmic rays~\cite{Stanev:2000fb}-\cite{Allard:2006mv}.
The need for detailed
simulation of photoproduction interactions in the energy range from
particle production threshold to the highest observed energies is
emphasized~\cite{Risse:2005gv}.\\

\noindent
In the second part of the presentation, currently available minimum
bias Monte Carlo codes for simulation of photoproduction are
reviewed, focussing on {\sc SOPHIA}~\cite{Muecke00a}, 
{\sc FLUKA}~\cite{Fasso01a}, {\sc RELDIS}~\cite{Pshenichnov:2001qd},
 and {\sc DPMJET III}~\cite{Roesler00a}.
Physics concepts applied for describing particle production
at different energies are introduced and their implementation in
simulation packages is discussed. Comparisons with measurements are
used to demonstrate the applicability of different programs. Finally
the limitations and the considerable uncertainties of simulating multiparticle
production simulations at very high energy are stressed.\\


%
\newabstract 

\begin{center}
{\large\bf Low-$x$ QCD via electromagnetic PbPb collisions at 5.5 TeV in CMS}\\[0.5cm]
{\bf David~d'Enterria}$^1$, {\bf Aurelien Hees}$^{1,}$\footnote{Currently at UC Louvain.} for the CMS collaboration\\[0.3cm]
$^1$CERN, Geneva, Switzerland\\
\end{center}

\noindent
Photoproduction of heavy vector-mesons in electromagnetic interactions (aka.
Ultra-Peripheral Collisions, UPCs) of heavy-ions at very high energies provides
direct information on the parton distribution function (PDF) in the nucleus
at low values of Bjorken-$x$~\cite{d'Enterria:2006ep}.
The capabilities for the measurement of $\ups$ produced in ultraperipheral
PbPb collisions at $\sqrtsnn$ = 5.5 TeV in the Compact Muon Solenoid
(CMS) experiment at LHC~\cite{PTDR1} are studied in the $\elel$ and $\mumu$
decay channels~\cite{d'Enterria:2007}.\\ 

\noindent
The measurement discussed here makes use of the CMS tracker, ECAL and 
muon chambers ($|\eta|<$ 2.5, full $\phi$) for muon and electron reconstruction, 
and the Zero Degree Calorimeters (ZDC)~\cite{Grachov:2006ke} for tagging the 
ultra-peripheral PbPb events accompanied by forward neutron emission.
The \STR event generator~\cite{starlight} is used as input Monte Carlo for the
double differential cross-sections of  (i) the $\ups$ photoproduction signal, 
and (ii) the most significant source of physical background: coherent production of 
high-mass lepton pairs in $\gaga\rightarrow\lele$ processes. The mass, $p_T$ 
and rapidity spectra of the $\ups\rightarrow \elel$,~$\mumu$ 
obtained after background subtraction with a full simulation and reconstruction analysis 
are presented. The excellent bottomonia mass resolution of the CMS muon chambers 
($\sigma_{\ups}\approx$~90~MeV/$c^2$) will allow 
one to measure separately the different vector states ($\ups,\ups',\ups''$). The dedicated
L1 and high-level triggers proposed for the UPC measurement are also discussed.
The final expected rates ($\sim$400 $\ups$'s) for a nominal PbPb run with integrated
luminosity of 0.5 nb$^{-1}$ will make possible a differential study of the
rapidity dependence of $\ups$ photo-production. Such a measurement is expected to 
constrain the underlying gluon PDF in the Pb nucleus in a kinematic regime
($x\approx$10$^{-3}$, $Q^2\approx$ 40 GeV$^2$) unexplored
so far~\cite{d'Enterria:2006nb}.\\

\noindent
{\bf Acknowledgments.}
\noindent
Work supported in part by the 6th EU Framework Programme under 
contract number MEIF-CT-2005-025073.

%
\newabstract 

\begin{center}
{\large\bf Trigger capabilities of ALICE TOF for Ultra-Peripheral Collisions}\\[0.5cm]
{\bf Eugenio Scapparone}$^{3}$ on behalf of the ALICE-TOF group\\[0.3cm]
$^1$ITEP Moscow, Moscow, Russia\\[0.2cm]
$^2$University of Bologna, Bologna, Italy\\[0.2cm]
$^3$INFN-Bologna, Bologna, Italy\\[0.2cm]
$^4$University and INFN of Salerno, Salerno, Italy\\[0.2cm]
$^5$Dept. of Physics, Kangnung National University, South Korea\\[0.2cm]
$^6$World Laboratory, Lausanne, Switzerland\\[0.6cm]
\end{center}

\noindent
The ALICE TOF detector is a natural candidate to address UPC selection.
The fast response of the MRPC, the large area covered at $|$$\eta$$|$$<$~1 and the high
segmentation make the TOF well suited for triggering at L0 level in the central region. 
Each Front End card (FEA) reads 24 MRPC pads and makes their OR. Such OR is daisy chained with the OR of the closer
FEA. The resulting signal (hereafter TOF-OR), made by the ORs of 48 pads,  ( i.e. $\simeq$ 500~$cm^{2}$)
is sent to the first TOF trigger layer, a VME board called Local Trigger Module (LTM). 
Each of the 72 TOF VME crates contains a LTM module, sending 24 bits in parallel to the 
Cosmic and Topology Trigger Module (CTTM), located 60 m far away, close to the ALICE CTP.
The CTTM is expected to take the L0 trigger decision and to send it to the CTP. 
UPC can produce lepton pairs in $\gaga$ interactions, and vector mesons and 
jets in photoproduction events.  
When triggering on the exclusive production of vector mesons, the most important parameter 
to face is the fake trigger rate (FTR), due to combinatorial background. Such events 
show up as few tracks in an otherwise empty detector. The key ingredient when computing 
the FTR, is the MRPC noise, whose measurement gives 
0.5 Hz/$cm^{2}$. Such noise is mainly due to ionizing particles in
the chamber. A guess on the noise that the TOF will experience when running in ALICE is
 not straightforward:  as a safety margin we 
will use for the prediction reported below a MRPC noise of 2.5 Hz/$cm^{2}$.
The FTR, when selecting a number of
fired $N_{TOF-OR}$~$\geq$~5 is $\simeq$ 1 Hz, while the FTR for 
$N_{TOF-OR}$~=~2 is 200~kHz. Such high rate, unmanageable also at L0 level, can be further reduced
by using the vector meson decay topology. We simulated the
J/$\Psi$ $\rightarrow$ $l^{+}l^{-}$ and the $\rho$$\rightarrow$$\pi\pi$
decay, using the \STR Monte Carlo. The particles
produced in the decay were traced in a empty cylinder in a B=0.5 T magnetic
field. We found an efficiency for containing both the decay products  
of $\epsilon_{cont}^{J/\Psi}$=16.7$\%$ and $\epsilon_{cont}^{\rho}$=8.3$\%$ respectively.
A full simulation is in progress. Taking advantage of the produced particle topology 
in the  J/$\Psi$ decay,  by selecting only the TOF-OR pairs in a 
$150^{o}$$\leq$$\Delta\phi$$\leq$$170^{o}$ window the
FTR can be reduced by a factor 18.  For the pions from $\rho$ decay by selecting only 
the TOF-OR pairs in a $70^{o}$$\leq$$\Delta\phi$$\leq$$110^{o}$ window, the FTR can be reduced
by a factor 9, while keeping 60$\%$ of the signal.
 A further FTR reduction can be obtained for both vector mesons decay
considering that in Pb-Pb interaction, despite a bunch crossing length of 125 ns,
the TTC will distribute a 40 MHz clock. 
Since the OR signal has a 20 ns length, we can align this signal
so that the positive edge of the TTC clock is well inside it. By enabling
the latching only in the edge effectively corresponding
to the bunch crossing, we can
reduce the noise by a factor of 5. In the combinatorial background, for 
$N_{TOF-OR}$ = 2, this corresponds to a factor 25, giving:\\\\
FTR$_{L0} < 200$ kHz/18/25 = 440 Hz~for $J/\Psi \rightarrow l^{+}l^{-}$ 
and~~FTR$_{L0}< 200$ kHz/9/25 = 880 Hz for $\rho\rightarrow\pi\pi$.\\\\
Such FTR at L0 level has to be compared with the genuine $\jpsi$ and $\rho$ rates:\\\\
Rate$_{J/\Psi}=L\cdot\sigma\cdot\epsilon_{cont}^{J/\Psi}\cdot\Gamma$=0.32 Hz ~~and~~
Rate$_{\rho}=L\cdot\sigma\cdot\epsilon_{cont}^{\rho}\cdot\epsilon_{\Phi}$=130 Hz.\\\\
The FTR$_{L0}$ rates can be reduced requiring a coincidence with L0
triggers from other central detectors.

%
\newabstract 

\begin{center}
{\large\bf Strong electromagnetic fields in heavy ion collisions:\\ multiphoton processes}\\[0.5cm]
{\bf Gerhard Baur}$^1$\\[0.3cm]
$^1$Forschungszentrum J\"ulich, Institut f\"ur Kernphysik, D-52425 J\"ulich, Germany\\
\end{center}

\noindent
The characteristic properties of ultraperipheral relativistic
heavy ion collisions are discussed. Very strong fields occur
for a very short time. This means that the corresponding
equivalent photon spectrum extends to energies which 
exceed present photon energies available at other facilities~\cite{rep}-\cite{arnps}.
Also, due to the high charge of the ions ($Z=79$ at RHIC
and $Z=82$ at the forthcoming LHC) multiphoton processes
occur in a single collision, see e.g.~\cite{npa}.\\

\noindent
One of the main interests is in 
coherent vector meson photoproduction ,
which was observed at RHIC and is relevant for low-$x$
QCD studies~\cite{thisworkshop}.
The transverse momentum distribution depends 
on an interference effect. The photon producing the 
vector meson can come from either ion ~\cite{ajk}.
The transverse momentum distribution
is determined by the transverse momentum distribution
of the photon and on the production process on the target,
this is studied in semiclassical and Glauber models in ~\cite{prl}.

%
\newabstract 

\begin{center}
{\large\bf A low multiplicity trigger for peripheral collisions in ALICE}\\[0.5cm]
{\bf Rainer Schicker}$^1$, on behalf of the ALICE Collaboration\\[0.3cm]
$^1$ Phys. Inst. University Heidelberg \\
\end{center}

\noindent
The ALICE experiment at the LHC is presently being built as a general 
purpose heavy ion detector. ALICE consists of a central barrel covering 
a pseudorapidity range $-0.9 < \eta < 0.9$ and a forward muon spectrometer 
covering a range $2.5 < \eta < 4.0$. Additional detectors 
are used for event characterization and for trigger purposes. 
  
\noindent
The physics programme of ALICE focuses on the study of strongly interacting 
matter at high energies with a multitude of experimental observables. 
Such a programme necessitates data taking in nucleus-nucleus as well as 
in nucleon-nucleon collisions. The total nucleon-nucleon cross section at LHC 
energies is expected to be about 100 mb with a substantial contribution 
of \mbox{30 mb} 
due to diffractive events. The subclass of central exclusive production 
is characterized by particle production at midrapidity and by the existence 
of a rapidity gap on either side. The kinematics of central exclusive 
production is matched to the acceptance of the ALICE detector, hence
measurements of some channels of diffractive central exclusive production 
become feasible.   

\noindent
In my talk I will present a trigger scheme for diffractive central exclusive 
events. I will describe how such a scheme can be realized within the existing 
ALICE detector systems. I discuss some physics channels which become accessible
with such a trigger and present some rate estimates.

%
\newabstract 

\begin{center}
{\large\bf NA60 experimental capabilities for ultra-peripheral InIn collisions}\\[0.5cm]
{\bf Carlos Louren\c{c}o}$^1$, on behalf of the NA60 Collaboration\\[0.3cm]
$^1$CERN-PH, Geneva, Switzerland\\
\end{center}

\noindent
In this presentation we introduce the NA60 experiment, in terms of
concept, detectors and performance, placing emphasis on issues of
relevance for the tagging of Ultra-Peripheral Collisions, which is
necessarily done in rather different ways in a fixed-target collision
geometry with respect to the presently more common collider
experiments. We also mention some specific features affecting the
NA60 case, such as beam and interaction pile-up, detector resolutions,
etc. In summary, it seems likely that the available NA60 In-In data
can provide useful information on dimuon production in
Ultra-Peripheral Collisions.

%
\newabstract 

\begin{center}
{\large\bf Dimuon production in Ultra-Peripheral In-In Collisions in NA60}\\[0.5cm]
{\bf Pedro Ramalhete }$^1$, on behalf of the NA60 Collaboration\\[0.3cm]
$^1$CERN, Geneva, Switzerland and IST, Lisbon, Portugal\\
\end{center}

\noindent
This presentation reports the status of the ongoing analysis to identify Ultra-Peripheral Collisions 
in the Indium-Indium data collected by NA60 in 2003.
A first iteration of the Event Selection procedure is described in detail, with emphasis on aspects 
that affect the NA60 data differently from what happens in collider experiments, such as beam 
and interaction pile-up.
We conclude that the use of several complementary NA60 detectors should provide a dimuon 
event sample where Ultra-Peripheral In-In Collisions have a very significant contribution.

%
\newabstract 

\begin{center}
{\large\bf UPC lepton pair production to all orders in $Z \alpha$ }\\[0.5cm]
{\bf Anthony J. Baltz}$^1$\\[0.3cm]
$^1$Brookhaven National Laboratory, Upton 11973, NY, USA\\
\end{center}

\noindent
Calculations of lepton pair production based on solution of the heavy ion
$\delta$ function potential Dirac equation are reviewed.
For the bound-electron positron problem the full solution of the
problem is in perturbation theory form, but with an eikonalized interaction
in the transverse direction.  This exact semiclassical solution produced a
reduction of a little less than 10\%  from perturbation theory in the predicted
cross section for Au + Au at RHIC.  One can identify this reduction as an
additional Coulomb correction from the moving ion not receiving the electron
to bound-electron positron pair production~\cite{1}.\\

\noindent
A two center light cone calculation of continuum pairs was made by solving the
semi-classical Dirac equation for colliding $\delta$ function
potentials~\cite{2}-\cite{4}.
Several authors have argued that a correct regularization of integrals
leading this exact
Dirac equation amplitude should lead to Coulomb corrections~\cite{5}-\cite{6}.
A physically motivated cutoff of the transverse potential effected
this correct regularization and leads to an
``exact'' cross section expression with Coulomb corrections~\cite{7}.
A full numerical evaluation of the ``exact'' total cross section for 
$e^+ e^-$ production with gold or lead ions shows reductions from perturbation
theory of 28\% (SPS), 17\% (RHIC), and 11\%(LHC), and
no final momentum region was found in which there was no reduction
or an insignificant reduction of the exact cross section~\cite{8}.
Reductions in the exact total probability of $e^+ e^-$ production from
perturbation theory were seen at all impact parameters~\cite{9}.
Suggested possibilities for observing Coulomb corrections at LHC were made:
(1) forward \boldmath $e^+ e^-$ pairs,
(2) $\mbox {\boldmath $\mu^+ \mu^-$}$ pairs,
and (3)deviations from $Z^4$ scaling.

%
\newabstract 

\begin{center}
{\large\bf High energy photon interactions at the LHC }\\[0.5cm]
{\bf K. Piotrzkowski}$^1$, on behalf of the UC Louvain Photon Group\\[0.3cm]
$^1$UC Louvain, B-1348 Louvain-la-Neuve, Belgium\\
\end{center}

\noindent
First collisions of 7 TeV protons at the CERN Large Hadron Collider are expected in 2008, 
when studies of interactions of proton constituents, quarks and gluons, at unprecedented 
energies will begin. However, protons are charged particles, and a significant fraction of pp 
collisions will involve high-energy interactions of photons exchanged by one, or both incoming 
protons. Most of the time such protons will stay intact and will be scattered at very small angles, 
and thanks to significant proton energy losses, tagging the high-energy photon-photon and 
photon-proton interactions by dedicated forward detectors becomes possible~\cite{kp}. 
Hence, by adding such detectors to the ATLAS and CMS experiments one can extend their 
physics reach and effectively convert the LHC into a high-energy photon-photon or a 
photon-proton collider. The same tagging technique can be used to 
select diffractive interactions at high luminosity~\cite{fp420}.\\

\noindent
The rate of the photon--induced electroweak processes is significant. Apart from the low-mass 
muon pairs, the highest cross-section of about 40 pb is obtained for the single $W$ boson photo-
production. One should note that a large cross-section of about 1 pb is expected for large 
photon-proton cms energies, $W_0 > 1$ TeV, so interesting studies will become possible already at 
initial low luminosity. It is therefore not surprising that the cross-section for the associated 
WH photoproduction is significant, above 20 fb for the SM light Higgs boson, and WH photoproduction 
constitutes more than 2\% of the total inclusive WH production at the LHC! Contrary to the pp case, 
the top pair photoproduction of 1.5 pb is not so overwhelming, and the top background will be much 
less severe, allowing for a complementary measurement of the WH production provided sufficient 
luminosity.\\

\noindent
Finally, the two-photon $W^+W^-$ exclusive production has the total cross-section of more than  
100 fb, and a very clear signature. Its cross-section is still about 10 fb for $W_0 >$  1 TeV showing 
sensitivity to physics beyond the SM. On the other hand, the two-photon dimuon production will be an 
excellent calibration process, with very well known cross-section from QED, and an extremely clear 
signature of the exclusive, back-to-back dimuons in the central detectors.\\

\noindent
These initial studies of photon induced high-energy interactions at the LHC show very 
interesting prospects. One should stress that apart from the spectacular exclusive muon pairs, 
all the other considered final states contain at least one $W$ boson. It means that even 
using the nominal LHC triggers one will be efficiently selecting also photon events. However, for the 
final selection, and in particular for the suppression of huge  inclusive pp backgrounds, tagging 
photon events by forward proton detectors is mandatory. In addition, it will improve the 
reconstruction of events by using the measured momenta of the forward-scattered protons. 
This will lead to very clean samples, in particular for the exclusive two-photon production �for 
example, for the fully leptonic decays of $W$ pairs the final state will consist of two forward protons, 
two very high pT central leptons of opposite sign, large missing energy and nothing else. 
Selection of such events should be therefore possible even at the nominal LHC luminosity.

%
\newabstract 

\begin{center}
{\large\bf  Anomalous single top photoproduction at the LHC}\\[0.5cm]
{\bf K. Piotrzkowski}$^1$ and {\bf J. de Favereau}$^1$ \\[0.3cm]
$^1$UC Louvain, Belgium, UE\\
\end{center}

\noindent
FCNC processes are expected in many extensions of the Standard Model. Measurements at HERA show that 
anomalous couplings $\kappa_{\gamma q t}$ can be strongly constrained by studying the single top 
photoproduction. Of course, in the Standard Model the single top cross-section is strongly suppressed, 
as it is zero at the tree level. The recent limits from the ZEUS and H1 collaborations~\cite{top-hera} 
significantly reduce the allowed range for anomalous couplings, in particular for $\kappa_{\gamma u t}$.
Rate of the photon-induced electroweak processes is significant at the LHC~\cite{kp}, and the energy 
reach is about 10 times higher than at HERA. Apart from the low-mass muon pairs, the highest 
cross-section of about 40 pb is expected for the single $W$ boson photoproduction, which is still at about 
1 pb for photon-parton cms energies above 1 TeV. In case of the direct photoproduction the $W$ boson is 
accompanied by at least one jet, making such W+jets events the major background in searches for the
anomalous single top photoproduction at the LHC, in case when the Wj invariant mass is close to the top 
mass and jet is misidentified as a b-quark jet. It should be noted that the NLO calculations has been
done for the HERA kinematics~\cite{spira}, showing complete dominance of the direct photoproduction 
for the $W$ boson transverse momenta of about 20 GeV, and above.\\

\noindent
Initial studies of such measurements at the LHC have been made using the adopted for photon 
interactions {\sc madgraph} package~\cite{fabio}, interfaced to {\sc pythia} for the final state hadronization, and
PGS - a generic, fast MC program to simulate detector and experimental effects~\cite{pgs}.
Leptonic decays of the $W$ boson, from the top quark decay, were assumed and b-jet tagging was requested.
The signal selection was done by requesting an electron or muon at large transverse momenta, significant
missing energy, the reconstructed invariant mass of the Wj system compatible with the top mass, and by a
cut on the direction of the b-jet in the c.m. frame of the Wj system. Photoproduction was identified or by
lack of significant energy flow into one of very forward calorimeters, or by requesting a hit in one of 
the roman pot proton detectors. Very preliminary results suggest that even at very low, initial LHC 
luminosity the present limits on $\kappa_{\gamma u t}$ and $\kappa_{\gamma c t}$ could be significantly 
improved. In the near future, the analysis will include full detector simulation and more complete 
background considerations.

%
\newabstract 

\begin{center}
{\large\bf Photoproduction and low-$x$}\\[0.5cm]
{\bf  M.V.T. Machado}$^1$, {\bf V.P. Gon\c{c}alves}$^2$\\[0.3cm]
$^1$Centro de Ci\^encias Exatas e Tecnol\'ogicas, Universidade Federal do Pampa, Bag\'e, RS, Brazil\\
$^2$ Instituto de F\'{\i}sica e Matem\'atica, Universidade Federal de Pelotas, Pelotas, RS, Brazil
\end{center}

\noindent
In this talk, we discuss photoproduction of heavy quarks (charm and bottom) and vector mesons 
($\rho$, $\phi$, $\omega$ and $J/\Psi$), which are directly accessible at DESY-HERA and in 
coherent processes at RHIC and the LHC. In the latter case, the high center of mass energy and the 
large luminosities produces high event rates.  The huge LHC energy will allow probing gluon 
distribution in nucleon and nuclei towards very low-$x$. In this region, saturation effects to the 
gluon distribution are expected to be increasing large.  In order to compute the cross section we 
use the successful color dipole approach, which gives a good description of current accelerator data 
on photoproduction. The dipole formalism is suitable for the implementation of saturation effects, 
which are introduced in the modeling of the dipole-target cross section. In addition, we take into 
account the nuclear shadowing effects. They are computed within the Glauber-Gribov formalism 
for inelastic shadowing.\\

\noindent
First, we consider the charm and bottom photoproduction, $\sigma\,(\gamma p \rightarrow Q\bar{Q}\,X)$, 
for DESY-HERA energy \cite{Goncalves:2003kp}. It is shown that the saturation models underestimate 
the experimental data at high energies and they provide a lower bound for the cross section. The total 
diffractive contribution is also calculated. In sequence, we compute the photonuclear cross section, 
$\sigma\,(\gamma A \rightarrow Q\bar{Q}\,X)$,  for different nuclei and show that the strength 
of nuclear effects is still low for heavy quarks \cite{Goncalves:2003zy}. Those cross section are input 
in coherent reactions in $pp$, $pA$ and $AA$ collisions. In that case, the production cross section is 
obtained by the convolution of the photon flux in the ion (or proton) with the corresponding cross section 
for photoproduction. The rates are high and the experimental signal is feasible. For details on calculations 
and results, see Refs. \cite{Goncalves:2003is}-\cite{Goncalves:2005ge}. In a second investigation, 
we consider the $\rho$ and $J/\Psi$ photoproduction, $\sigma\,(\gamma p \rightarrow V_M\,X)$, 
which is compared with DESY-HERA data \cite{Goncalves:2004bp}. Then, we make prediction for the 
photonuclear cross section, $\sigma\,(\gamma A \rightarrow V_M\,X)$, for different nuclei and energies 
to be accessible at LHC. The color dipole approach gives a unified framework to compute cross sections 
for both light and heavy mesons. Once again, the rates are large and the typical topology of the final states 
makes the signal clear \cite{Goncalves:2005ge}-\cite{Goncalves:2005yr}. In both cases, the total inclusive cross 
section is shown to be smaller than the usual collinear perturbative QCD calculations.

%
\newabstract 

\begin{center}
{\large\bf Coulomb Corrections in the Lepton-Pair Production in Ultrarelativistic Nuclear Collisions }\\[0.5cm]
{\bf Mehmet Cem G\"{u}\c{c}l\"{u}}$^1$, {\bf Melek Y\i lmaz \c{S}eng\"{u}l}$^{1,2}$\\[0.3cm]
$^1$ \.{I}stanbul Technical University, Physics Department, Maslak- \.{I}stanbul, Turkey\\
$^{2}$ Kadir Has University, Faculty of Arts and Science, Cibali- \.{I}stanbul, Turkey\\
\end{center}

\noindent
The strong electromagnetic field of heavy ions can produce different kind of two-photon 
reactions at relativistic heavy ion colliders. Recently the  STAR collaboration has 
presented~\cite{adams} results on electron-positron pair productions on ultra-relativistic 
peripheral collisions. The authors compare the experimental data with QED and equivalent 
photon approximation (EPA).\\

\noindent
We have calculated the electron-positron pair production cross section by using second-order 
Feynman diagrams. We have employed Monte Carlo methods and solved it exactly. We have 
generalized this calculation for all energies and charges of the heavy ions. This gives us a 
semi-analytic cross section and impact parameter dependence of cross-section expressions. 
Monte Carlo calculation gives an equation for the impact parameter dependence cross section 
valid for all impact parameters. In these impact parameter regions, the electromagnetic field 
is very high and a detailed study of this region is important for nonperturbative effects.
We then compare our results with Born methods that include Coulomb corrections. The 
method loses applicability at impact parameters less than the Compton wavelength of the 
electron, which is the region of greatest interest for the study of nonperturbative effects. 
We present our results and argue that the Coulomb correction terms are not exact and these 
terms need to be improved.
On the other hand, the authors of Ref.~\cite{lee} made a small-momentum approximation  
and obtained an analytic expression. Although most of the integration comes from the 
small-momentum range, the lowest order in transverse momentum is not adequate to 
obtain accurate Coulomb corrections and higher orders should be also included. This was 
first noticed by Baltz~\cite{baltz}, and in this work we were also convinced that the 
small-momentum approximation  alone is not adequate to obtain correct Coulomb corrections.\\

\noindent
Recent publications about peripheral relativistic heavy ion collisions~\cite{bertulani}-\cite{hencken} 
show that the impact parameter dependence cross sections of lepton-pair production are very 
important and detailed knowledge of impact parameter dependence cross sections particularly 
for small impact parameters can help to understand many physical events in STAR experiments.

%
\newabstract 

\begin{center}
{\large\bf Photon-Photon and Photon-Nucleus Collisions\\ in ALICE at the LHC }\\[0.5cm]
{\bf Joakim Nystrand}$^1$, on behalf of the  ALICE Collaboration\\[0.3cm]
$^1$Department of Physics and Technology, University of Bergen, Bergen, Norway\\
\end{center}

\noindent
ALICE is a general purpose heavy-ion experiment at the Large Hadron Collider 
(LHC) at CERN. Its primary aim is to study central nucleus-nucleus 
collisions (Pb+Pb) in order to investigate the properties of highly 
excited nuclear matter and the quark-gluon plasma~\cite{Alessandro:2006yt}.
This talk deals with electromagnetic (photon-induced) interactions in 
ultra-peripheral collisions in which there is no overlap between the 
colliding ions~\cite{Bertulani:2005ru}. These events will have a very 
different topology compared with the central, hadronic interactions. 
They will therefore require different trigger and analysis techniques but 
will also broaden the physics potential of ALICE.\\ 

\noindent
ALICE is designed to handle charged particle multiplicities up to 
$d n_{ch} / d \eta =$ 8000, far above what is expected for ultra-peripheral 
collisions. Reconstructing the ultra-peripheral events should thus 
not be a problem. 
The major challenge will instead be triggering without a prohibitively high 
background rate. 
The Time-of-Flight and Si-pixel detectors can provide a 
low-multiplicity trigger at mid-rapidity. The foreseen ultra-peripheral 
trigger will combine the information from these detectors with a requirement 
of no signal 
in the V0 and T0 detectors. These are scintillator (V0) and 
Cherenkov counters (T0) located at forward rapidities on both sides 
of the collisions point with full azimuthal coverage for  
pseudo-rapidities $\approx 2 \leq | \eta | \leq 5$. \\

\noindent
The expected rates for several interesting ultra-peripheral interactions 
within the ALICE acceptance are high. For coherent
production of the heavy vector mesons $J / \Psi$ and $\Upsilon$, rates of 
150,000 and 600, respectively, in the $e^+ e^-$ decay channel can be 
expected within one ALICE year (corresponding to $10^6$~s). 
The rates for photoproduction of heavy-quarks and jets will also be 
high, and the yield from these processes within the ALICE acceptance 
should be significant.

}

\end{document}